\documentclass[12pt,preprint]{aastex}
\usepackage{graphics}

\begin{document}
\title{The Konus-Wind and Helicon-Coronas-F detection of the giant $\gamma$-ray
flare from the soft $\gamma$-ray repeater SGR 1806-20}

\author{E. P. Mazets\altaffilmark{1}, %
    T. L. Cline\altaffilmark{2}, %
    R. L. Aptekar\altaffilmark{1}, %
    D. D. Frederiks\altaffilmark{1}, %
    S. V. Golenetskii\altaffilmark{1}, \\%
    V. N. Il'inskii\altaffilmark{1}, %
    \& V. D. Pal'shin\altaffilmark{1}}

\altaffiltext{1}{Ioffe Physico-Technical Institute, St Petersburg 194021, Russia}
\altaffiltext{2}{NASA Goddard Space Flight Center, Code 661, Greenbelt, MD 20771}

\begin{abstract}
The giant outburst from SGR 1806-20 was observed on 2004 December 27
by many spacecraft (ref.~1,2,3,4,5,6). This extremely rare event
exhibits a striking similarity to the two giant outbursts thus far
observed, on 1979 March 5 from SGR 0526-66 (ref.~7) and 1998 August
27 from SGR 1900+14 (ref.~8,9,10). All the three outbursts start
with a short giant radiation pulse followed by a weaker tail. The
tail pulsates with the period of neutron star rotation of $\sim$5--8
s, to decay finally in a few minutes. The enormous intensity of the
initial pulse proved to be far above the saturation level of the
gamma-ray detectors, with the result that the most valuable data on
the time structure and energy spectrum of the pulse is lost. At the
time of the December 27 outburst, a Russian spacecraft Coronas-F
with a $\gamma$-ray spectrometer aboard was occulted by the Earth
and could not see the outburst. It succeeded, however, in observing
a weak reflected signal due to the $\gamma$-rays Compton scattered
by the Moon (ref.~11). This has been the first observation of a
cosmic gamma-ray flare reflected from a celestial body. Here we
report, that the detection of a weakened back-scattered initial
pulse combined with direct observations by the Konus $\gamma$-ray
spectrometer on the Wind spacecraft permitted us to reliably
reconstruct the intensity, time history, and energy spectra of the
outburst.
\end{abstract}

\section*{}
\clearpage
  During the period elapsed since its launch in 1994, Konus-Wind has detected
three giant outbursts. The first of them came on 1998 June 18 from SGR 1627-41
(ref.~12). It was weaker than the others and did not have a pulsating tail. The second giant outburst, on 1998 August 27 (ref.~8,9,10), was associated with SGR 1900+14.
It is to this event that the giant outburst on 2004 December 27 from SGR 1806-20 is very
similar.

\begin{figure}
\includegraphics[bb=60 180 490 730]{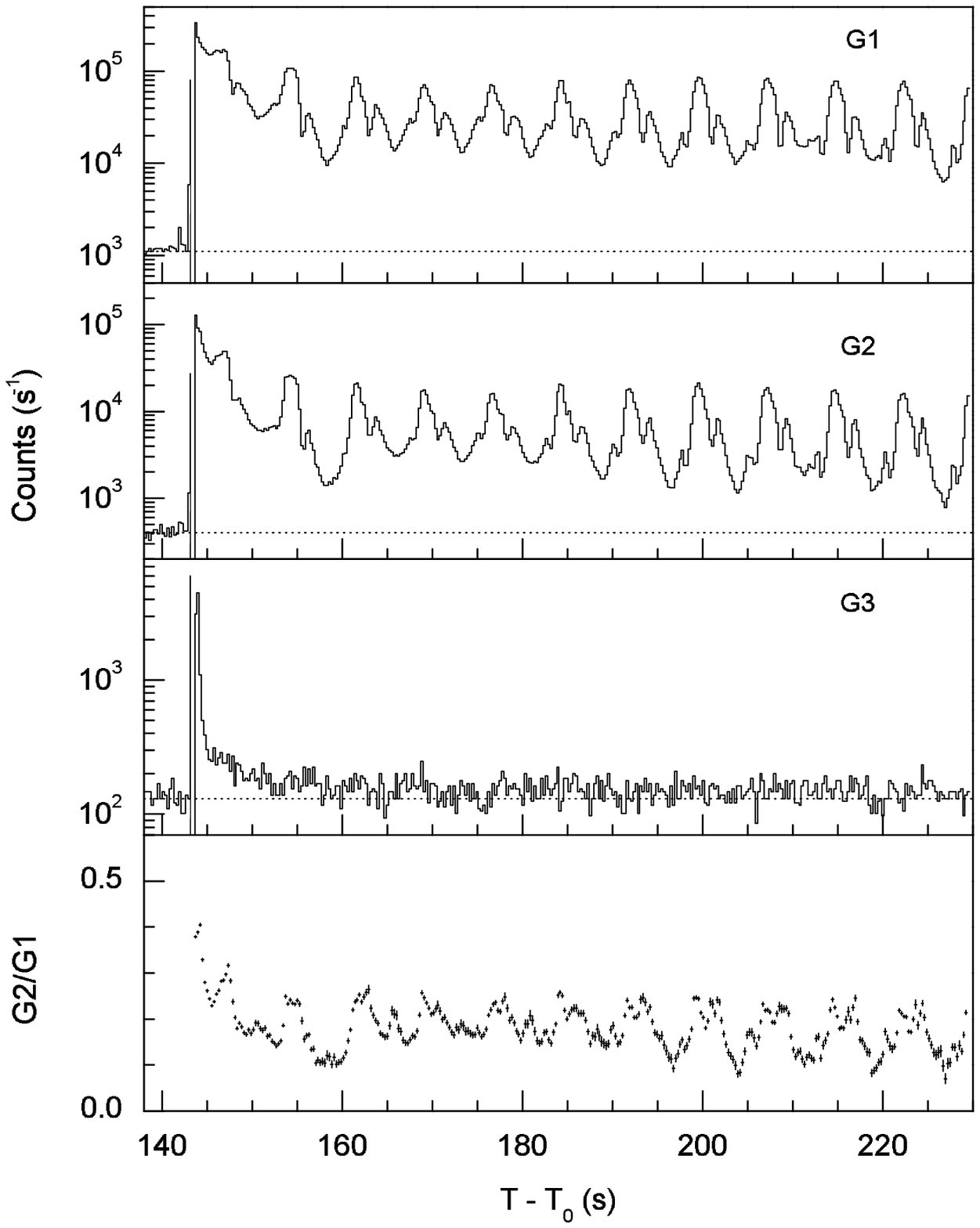}
\figcaption{Time history of the 2004 December 27 giant outburst
recorded by the Konus-Wind detector in three energy windows G1
(16.5--65~keV), G2 (65--280~keV), and G3 (280--1060~keV), and the
hardness ratio G2/G1. The moderate initial count rate growth to
10$^2$--10$^3$~counts~s$^{-1}$ transforms rapidly to an
avalanche-type rise to levels $>5 \times 10^7$~counts~s$^{-1}$,
which drives the detector to deep saturation for a time $\Delta T
\simeq 0.5$~s. After the initial pulse intensity has dropped to
$\sim 10^6$~counts~s$^{-1}$, the detector resumes operation to
record the burst tail. \label{Wind_TH}}
\end{figure}

\begin{figure}
\includegraphics[bb=60 180 490 730]{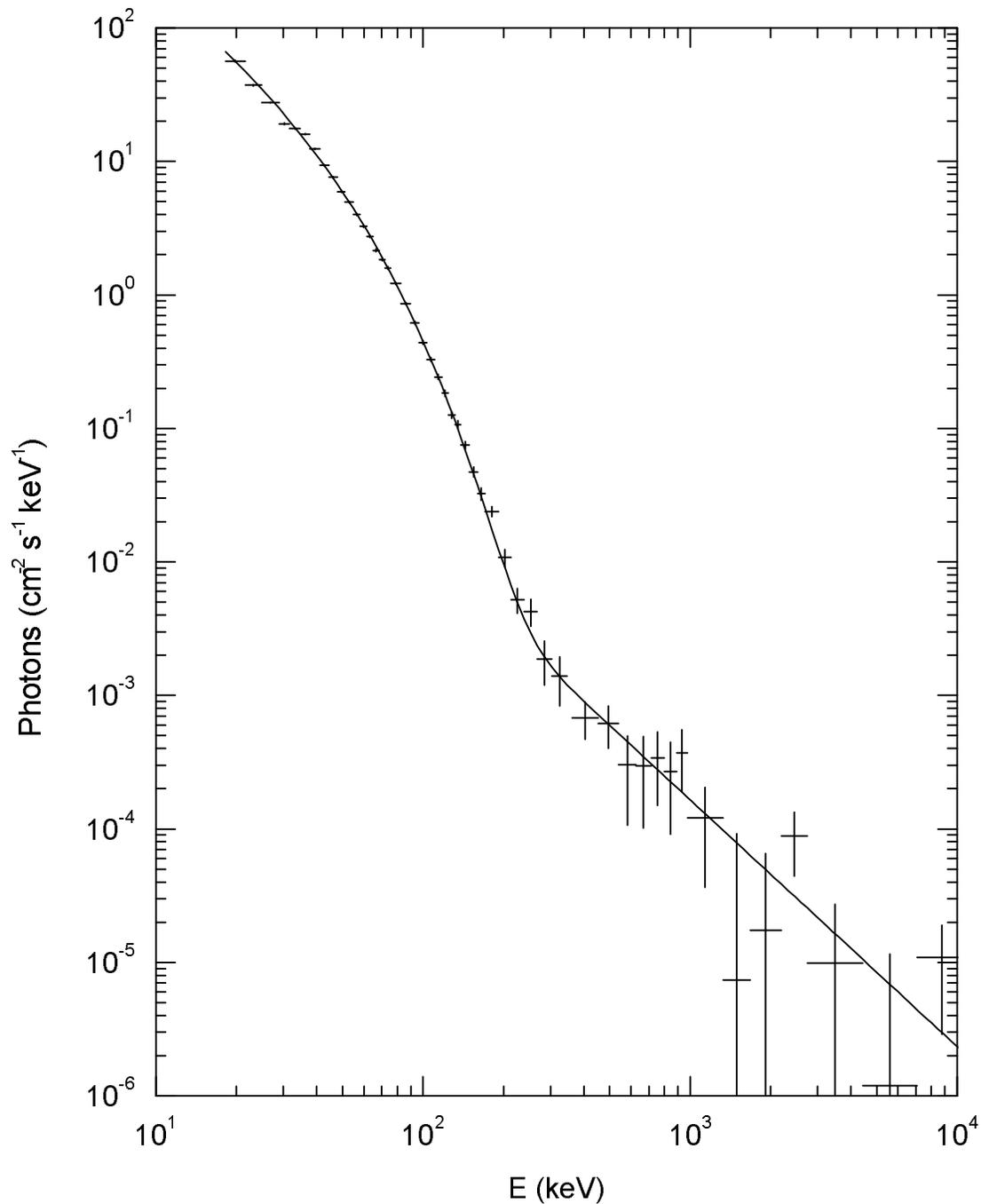}
\figcaption{A spectrum of the burst tail averaged over the pulsation
period. The low-energy component is similar to spectra of SGR's
recurrent bursts with $E_0 \simeq 30$~keV. At high energies it
exhibits a hard power-law component with $\alpha = -1.8 \pm 0.2$.
This two-component model is shown by the solid line.
\label{Wind_Sp}}
\end{figure}

     On Wind, the record of this outburst was triggered by a preceding recurrent soft
burst at $T_0$=21:27:58.447~s~UT. This burst was the last but
strongest of a series of 10 recurrent bursts which occurred on
December 27 before the giant outburst. The strongly enhanced SGR
recurrent activity at the end of 2004 could be an indication of an
approaching giant outburst, as was the case with the 1998 August 27
event (ref.~13). Figure~\ref{Wind_TH} presents a fragment of the
time history of the giant outburst recorded with a 0.256-s
resolution. The event started at $T - T_0$=142.08~s with an
extremely steep intensity rise to drive the gamma detector far above
the saturation level for $\sim$0.5~s. On termination of the initial
pulse and with the detector having resumed operation, the burst tail
became visible until its decay after $\sim$380~s. The burst tail
intensity pulsates with the period of neutron star rotation $P=7.57
\pm 0.07$~s. Each period exhibits a complex, three-peaked pulsation
structure. The count rate ratio of the G2 (65--280~keV) to G1
(16.5--65~keV) energy windows is a measure of the spectral hardness
of the radiation. Within each period, the spectral hardness
correlates with the intensity structure. Most likely the variations
in the intensity and spectral shape during a period are caused by
changing of the viewing angle on a rotating anisotropic source.
Figure~\ref{Wind_Sp} displays an energy spectrum of the tail
averaged over the pulsation period. The spectrum consists of two
components, with the energy region of up to $\simeq$400~keV being
dominated by a component falling off exponentially with energy, and
at energies $>400$~keV by a power law with a photon index of $-1.8
\pm 0.2$. Such spectra vary very little in shape as the burst
decays. Each of them contains a weak hard power-law component
extending to as far as 10~MeV. The total fluence in the outburst
tail can be estimated as $1.2 \times 10^{-2}$~erg~cm$^{-2}$ in the
20~keV--10~MeV band. Estimation of the initial pulse fluence from
complete detector saturation during $\approx$0.5~s can yield only
its lower limit $>10^{-2}$~erg~cm$^{-2}$. Thus, direct observation
of an outburst does not permit one to obtain data on the intensity,
time history, and energy spectrum of the initial pulse.

\begin{figure}
\includegraphics[bb=60 300 490 730]{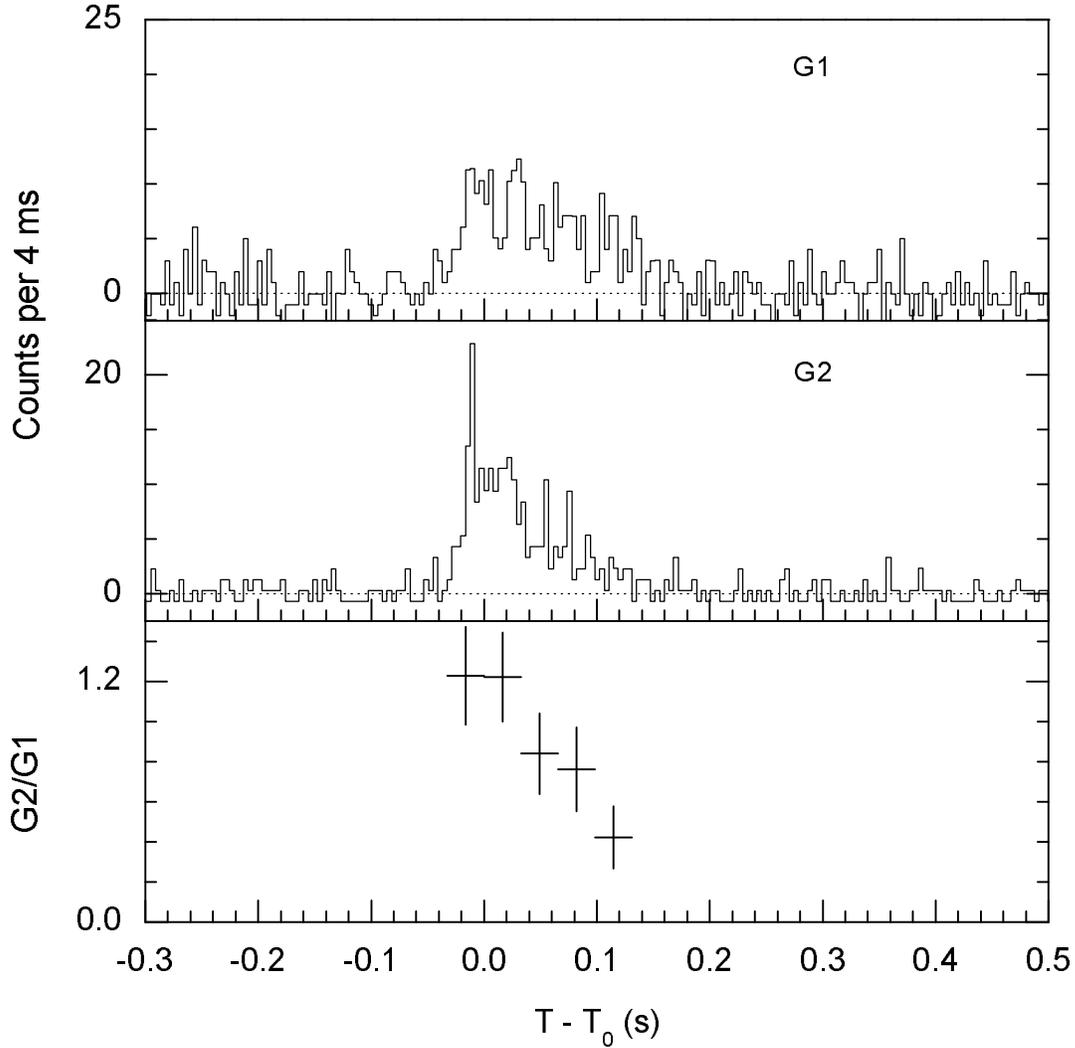}
\figcaption{Time history of the initial pulse of the outburst
reflected from the Moon as recorded by Helicon-Coronas-F
spectrometer. Time profile of the reflected pulse $F(t)$ is seen
during $\sim 200$~ms. Compared to the true profile of the initial
pulse $I(t)$, it is spread, and the rise is less steep because of
the signal being delayed in reflection from different regions on the
spherical Moon. If $t$ is measured in units of $R_M / c$, where
$R_M$ is the Moon's radius, and $c$ is the velocity of light,
$F(t)=\int_0^2 I (t - \tau) (1 - \tau/2) d \tau$.  \label{Hel_TH}}
\end{figure}

\begin{figure}
\includegraphics[bb=60 180 490 730]{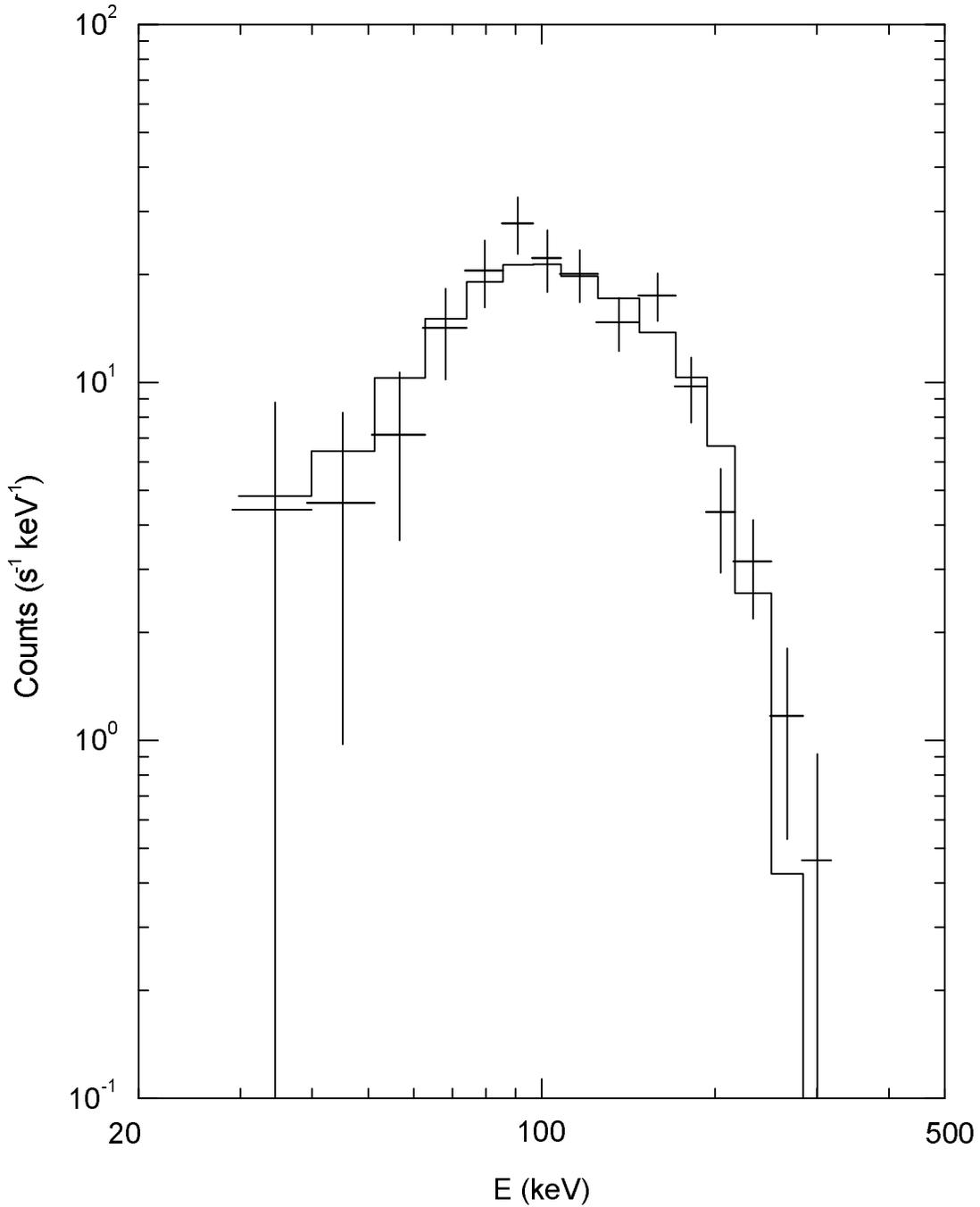}
\figcaption{Energy spectrum of the reflected signal. The
experimental points are shown with $±1 \sigma$ error bars. The
histogram is the initial pulse spectrum in form of
$dN/dE=A(E/100)^\alpha \exp(-E/E_0)$ modified by scattering from the
lunar soil. The best-fit parameters are: $A = 1.55 \times
10^4$~photons~cm$^{-2}$~s$^{-1}$~keV$^{-1}$, $\alpha = -0.7$, $E_0 =
800$~keV ($\chi^2 / d.o.f = 10.2/15$) \label{Hel_Sp}.}
\end{figure}

\begin{figure}
\includegraphics[bb=40 430 540 730, width = \textwidth]{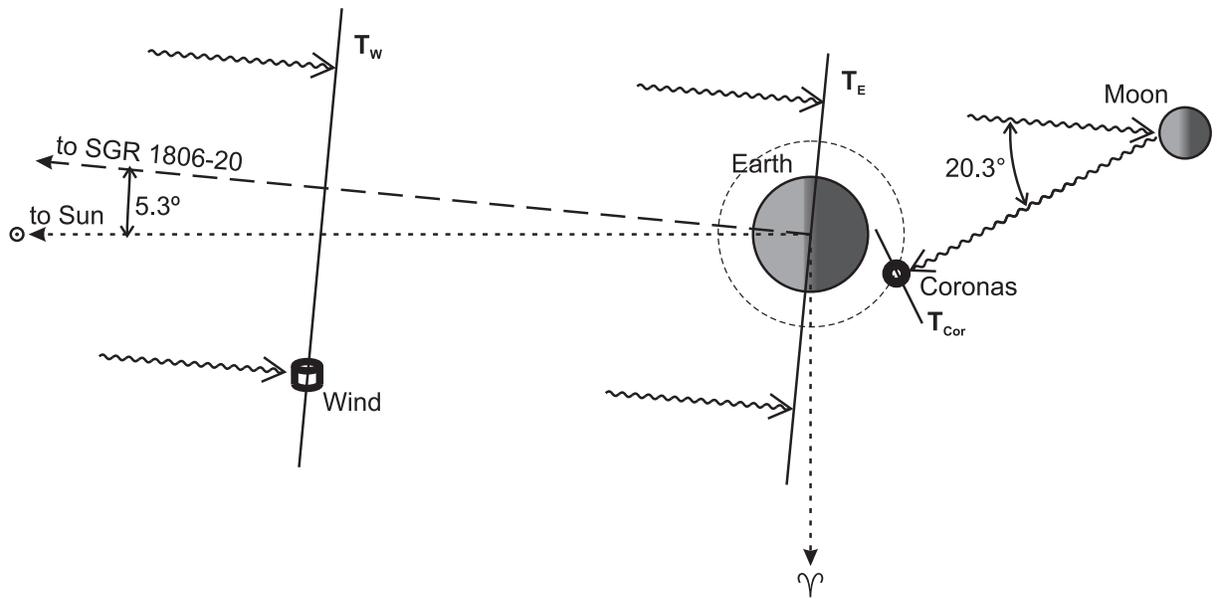}
\figcaption{A schematic diagram of the flare detection by the
Konus-Wind and Helicon-Coronas-F instruments. The flare wavefront
came from the SGR 1806-20 direction, crossed Wind at $T_W$, passed
Earth at $T_E =T_W + 5.086$~s, reached and was reflected by the
Moon, and, finally, back-scattered radiation was detected by
Helicon-Coronas-F at $T_{Cor} = T_W + 7.69$~s. \label{Schema}}
\end{figure}

     A fortunate chance to gain this information has been provided by observation by
the Helicon instrument on Coronas-F of a short gamma-ray burst on
2004 December 27 at 21:30:29.303~s~UT. Figure~\ref{Hel_TH} presents
a time history of this burst recorded in the energy windows G1
(25--100~keV) and G2 (100--450~keV), and the G2/G1 ratio which
exhibits a strong hard-to-soft spectral evolution. Shown in
Fig.~\ref{Hel_Sp} is an average energy spectrum accumulated over
128~ms after $T_0$. There are two grounds arguing for this burst to
result from Compton scattering of the giant pulse by the Moon.
First, it is consistent with the delay of its arrival to Coronas-F
and Wind to be expected in this case (a schematic diagram of the
flare wavefront propagation is shown on Fig.~\ref{Schema}). Second,
while the shape of the measured energy spectrum is far from typical
of gamma-ray bursts, it fits quite well the so-called Compton
backscattering peak at scattering angles of about 180$^\circ$, where
the energy of singly scattered photons cannot exceed $\sim mc^2 /
2$. The real angle through which the radiation incident on the
Coronas-F detector was scattered lies in the range
159.5--159.9$^\circ$. The right-hand wing of the observed spectrum
derives primarily from scattering, multiple as well as single, of
photons with energies in excess of $\sim$200~keV striking the Moon.
In this energy region, the cross section of photoabsorption is
substantially lower than that of Compton scattering. The left-hand
wing covers a low energy region where photoabsorption becomes
strongly dominating over scattering, so that an ever decreasing
fraction of photons is capable of escaping out of the lunar soil.

\begin{figure}
\includegraphics[bb=60 370 520 720]{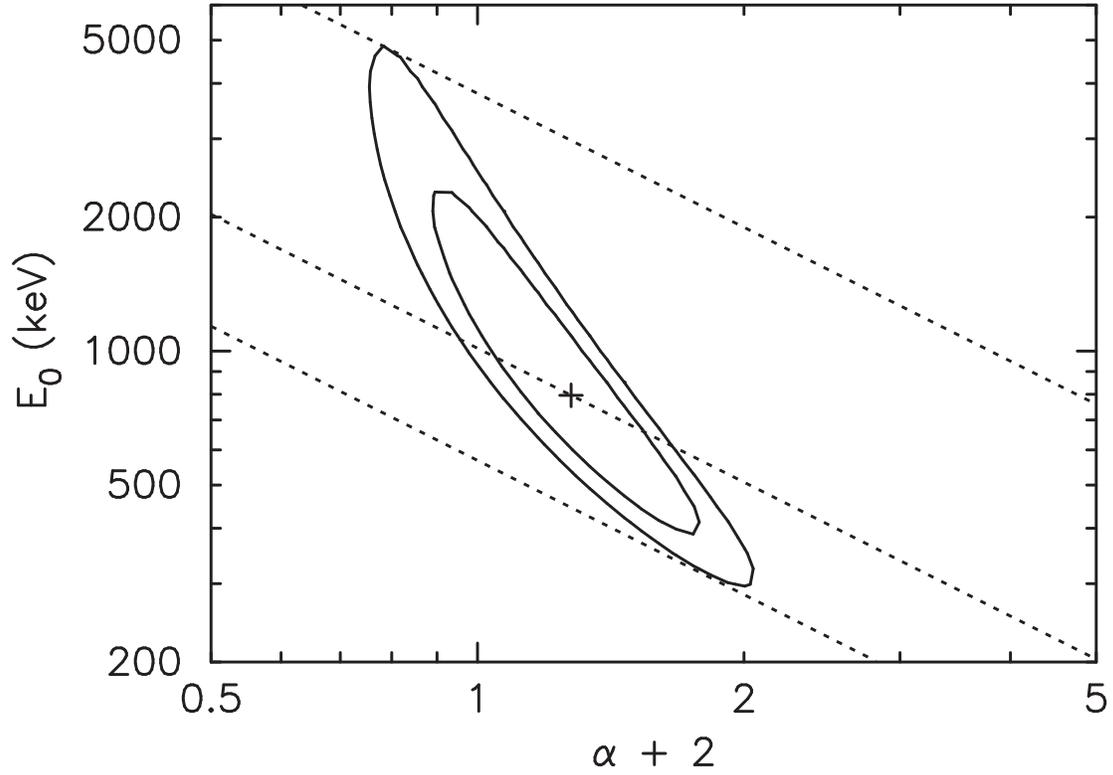}
\figcaption{Contour plot for $E_0$ vs $(\alpha + 2)$ derived from
the energy spectrum of the reflected signal. The two contours
correspond to 68 and 90\% confidence levels; the cross is relative
to the best fit values of the two parameters. $E_{peak} =(\alpha +
2)E_0$ is the energy of the maximum in a $\nu F_{\nu}$ spectrum.
Dependencies $E_{peak} = Const$ are also shown by dashed lines for
$E_{peak}$=0.6, 1.0, and 3.7~MeV. \label{Contour}}
\end{figure}

     To determine the spectrum and intensity of the initial pulse, the burst radiation
scattering from the Moon was numerically simulated by the Monte
Carlo method using the well-known GEANT code developed in CERN
(ref.~14). A response matrix of the Moon for the above-mentioned
scattering angles was calculated for a broad photon energy range
from 20~keV to 12~MeV and folded then with the detector response
matrix. The result of the spectrum simulation is also shown in
Fig.~\ref{Hel_Sp}. The best-fit model was found to be a power law
function with an exponential cutoff, $dN/dE = A \, (E/100)^\alpha \,
\exp(-E/E_0)$ with $\alpha$=-0.7 and $E_0$=800~keV. The reflected
energy flux seen by Coronas-F is weak being attenuated by a factor
of $\approx 10^{-6}$. The energy spectrum statistics is poor,
particularly in the wings. Therefore, the confidence regions of 68
and 90\% for the $\alpha$ and $E_0$ parameters shown in
Fig.~\ref{Contour} are fairly large. One readily sees, however, that
the fitting parameters $\alpha$ and $E_0$ are in strong
anticorrelation. As a result, the estimates of the total energy in
the pulse vary rather weakly. At a confidence level of 90\% the
fluence and the peak flux of the initial pulse in the 20~keV--10~MeV
energy band are $0.61^{+0.35}_{-0.17}$~erg~cm$^{-2}$ and
$9.2^{+5.6}_{-3.1}$~erg~cm$^{-2}$~s$^{-1}$.

\begin{figure}
\includegraphics[bb=70 180 500 720]{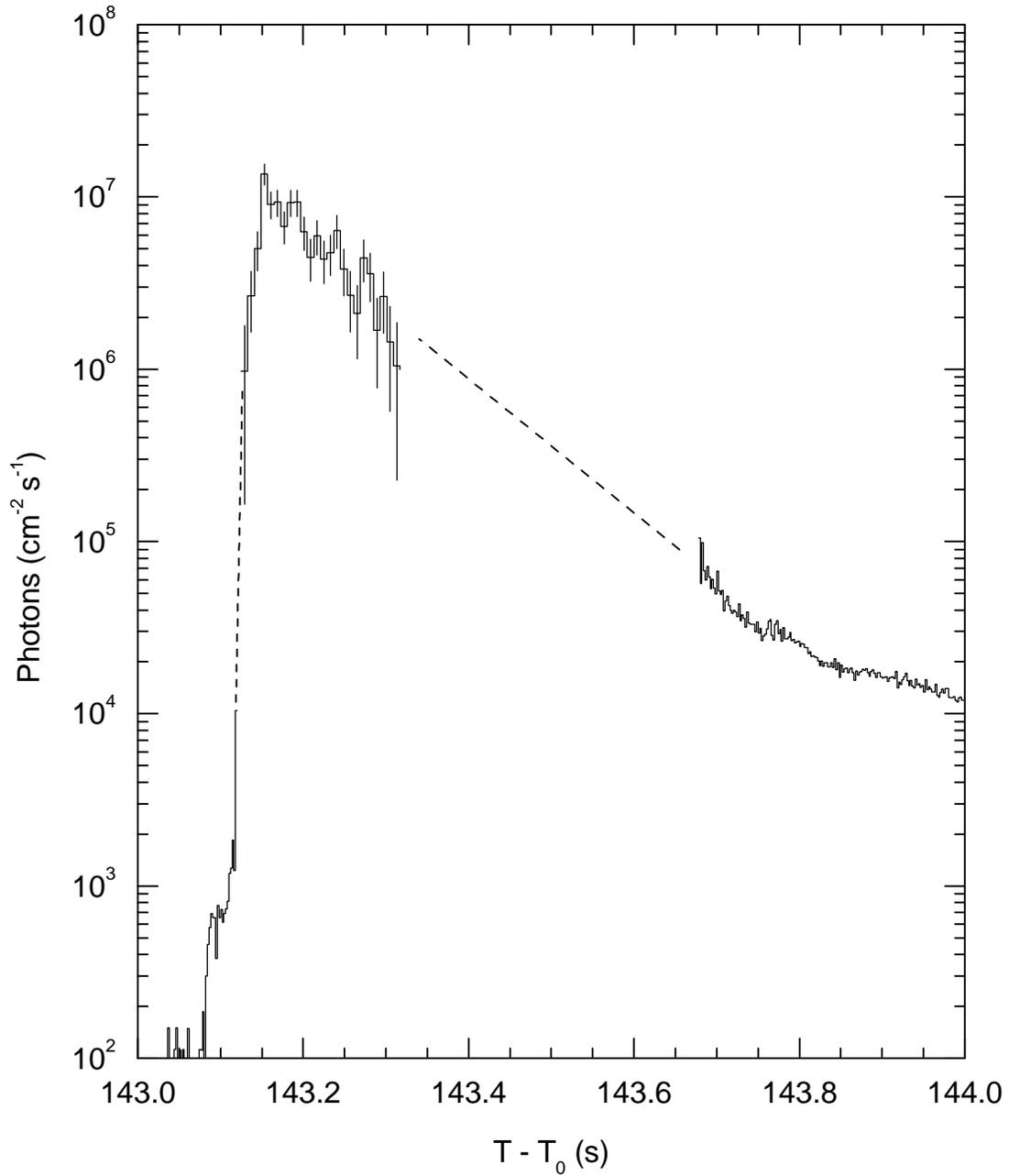}
\figcaption{Reconstructed time history of the initial pulse. The
upper part of the graph is derived from Helicon data while the lower
part represents the Konus-Wind data. The dashed lines indicate
intervals where the outburst intensity still saturates the
Konus-Wind detector, but is not high enough to be seen by the
Helicon. \label{ReconstrTH}}
\end{figure}

     Photons scattered from different zones of a spherical target reach an observer with
a relative time delay whose maximum value is $2R_M/c = 11.6$~ms,
where $R_M$ is the Moon's radius, and $c$ is the velocity of light.
This gives rise to a blurred time profile of the reflected signal
(Fig.~\ref{Hel_TH}). Correspondingly, the real front of the initial
pulse should be shorter than observed, $\approx$10--15~ms.
Figure~\ref{ReconstrTH} displays a reconstructed time profile of the
initial pulse. By Wind's information, the giant outburst starts with
a comparatively slow intensity rise, which transforms in $\sim$20~ms
to an avalanche-type growth. The continuation of the time history
was derived from Coronas-F data by introducing the corresponding
corrections for attenuation and spreading of the signal as it is
scattered and propagates to Coronas-F. As evident from the figure,
the intensity reaches a peak of $\sim
10^7$~photons~cm$^{-2}$~s$^{-1}$. At Helicon sensitivity, the
initial pulse is detected over a background for $\sim$150~ms. A
saturation of the Konus lasts longer, about 600~ms.

     Assuming isotropic emission and a distance to SGR 1806-20 of 15~kpc (ref.~15),
the energy release and the maximal luminosity of the initial pulse
are $1.6 \times 10^{46}$~erg and $2.5\times10^{47}$~erg~s$^{-1}$.
The energy release in the tail of all the three giant outbursts in
the SGRs is $\sim 10^{44}$~erg, and one may expect the energy
confined in their initial pulses to be comparable, about $\sim
10^{45}-10^{46}$~erg. This is consistent with our earlier assumption
that the lower limit on the energy released in the initial pulse of
the giant outbursts from SGR 1900+14 is actually many times smaller
than its true value (ref.~9), because the referred value of
$>7\times 10^{43}$~erg was calculated for a minimum flux which is
capable of saturating the Konus. The energy release of 10$^{46}$~erg
is comparable to the energy stored in the magnetosphere of a neutron
star with a surface magnetic dipole field of $\sim 10^{14}$~G. This
may present serious difficulties for the magnetar model of SGRs
(ref.~16). Giant outbursts from SGRs at larger distances should be
observed as short ($\sim$0.25 s) gamma-ray bursts with a hard
spectrum (ref.~17). It is conceivable that it is such events that
make up part of the Konus-Wind catalog of short gamma-ray bursts
(ref.~18). Present-day gamma-ray spectrometers, which can reliably
measure the time profile and energy spectrum of short gamma-ray
bursts with fluences of $\sim 10^{-7}$~erg~cm$^{-2}$, should be
capable of recording the initial pulses of giant outbursts from SGRs
at distances of $\sim$30~Mpc. If such a burst is localized to within
$\sim$5--10~arc~minutes, its identification with a host galaxy may
become possible.

\clearpage

\end{document}